\documentclass[preprint, 12pt,nofootinbib]{revtex4-1}

\usepackage{graphicx}
\usepackage{cancel}
\usepackage{amssymb}
\usepackage{textcomp}
\usepackage{amsmath}
\usepackage{bm}
\usepackage{times}
\usepackage{epsfig}
\usepackage{color}
\def\f{\frac}
\def\lf{\left}
\def\rh{\right}

\def\beq{\begin{equation}}
\def\eeq{\end{equation}}
\def\be{\begin{equation}}
\def\ee{\end{equation}}
\def\bea{\begin{eqnarray}}
\def\eea{\end{eqnarray}}
\def\to{\rightarrow}

\begin{document}
\title{Higgs search and flavor-safe fermion mass generation}
\author{Hui Luo}
\email{huiluo@zimp.zju.edu.cn}
\author{Ming-xing Luo}
\email{luo@zimp.zju.edu.cn}
\author{Kai Wang}
\email{wangkai1@zju.edu.cn}
\affiliation{Zhejiang Institute for Modern Physics and Department of Physics, Zhejiang University, Hangzhou, Zhejiang 310027, CHINA}

\begin{abstract}
We study a scenario of electroweak symmetry breaking where the weak gauge boson masses arises significantly from a fermiophobic source.  
To minimize flavor violation, the fermion mass generation is still due to one light doublet scalar. One of the realizations is the Bosonic
Technicolor model. In these scenarios, the Yukawa couplings between the light scalar and the standard model fermions are in general enhanced while the couplings between the light scalar and weak gauge bosons are reduced. Even though the flavor violation induced by the 
neutral scalar at the tree level can be avoided, 
the charged scalar state inevitably mediate flavor changing
neutral current processes. With the enhancement in the Yukawa
couplings, one expects serious constraints of such models from flavor
violating effects. We find that the most severe bound comes 
from neutral meson  mixing of  $B^{0}_{d}-\overline{B^{0}_{d}}$. 
Large parameter space is excluded if the weak gauge boson 
mass generation is dominated by the fermophobic sector. 
However, the correlation between the Yukawa coupling 
and charged scalar mass show that a factor of two 
enhancement in top Yukawa coupling is still allowed
for charged scalar heavier than 500~GeV. 
We use this as a benchmark point to study the phenomenology
of the light scalar. It is interesting that the destructive interference
between the top quark loop and the $W$-boson loop in the di-photon channel becomes significant and makes the channel negligible. 
In the light scalar region, the search becomes much more 
challenging than the conventional SM Higgs boson. 
\end{abstract}

\maketitle

\section{Introduction}

The physics mechanism of spontaneous electroweak symmetry breaking
(EWSB) remains an open question in the standard model (SM)
and there are good reasons to expect it to be unveiled by direct evidence from current collider experiments.  The simplest and the most economic mechanism of EWSB is 
the minimal Higgs boson model where one fundamental scalar
$SU(2)_{L}$ doublet develops a vacuum expectation value ({\it vev})
that gives rise to weak gauge boson masses through gauge interactions and SM fermion masses through Yukawa interactions.  
At hadron colliders, the search of SM Higgs of extensive mass range 
relies on the fact that the Higgs boson couples
to both weak gauge bosons and fermions simultaneously.
For instance, for $m_{h}\lesssim 140$~GeV, 
the leading discovery channel of the Higgs boson is through $W^{\pm}h$ or $Zh$ associated
production with $h\to b\bar{b}$ and the gluon fusion Higgs production $gg\to h$ with
$h\to \gamma\gamma$ where production is only due to Yukawa couplings to heavy quarks
while the decay is dominated by the $W$-loop. An alternative channel
is the weak boson fusion production with $h\to \tau^{+}\tau^{-}$.
For heavy SM Higgs with $m_{h}>140$~GeV, $gg\to h$ with $h\to W^{+}W^{-}$ ($WW^{*}$)
or $ZZ$ becomes the leading discovery channels.
These channels play important role not only in discovering the Higgs boson but also in confirming the role of Higgs in EWSB.

Both electroweak gauge boson masses and the SM fermion masses break
electroweak symmetry. However, since the gauge kinematic terms are invariant under the chiral transformation,
 the generation of SM fermion masses requires additional
breaking of chiral symmetries
$U(3)_{Q}\otimes U(3)_{u}\otimes U(3)_{d}\otimes U(3)_{\ell}\otimes U(3)_{e}$
even after the EWSB takes place. On the other hand,
in the absence of the SM Higgs, the scattering amplitudes for the longitudinally polarized gauge bosons ($W_{L}$ and $Z_{L}$) grow with energy squared $E^{2}$  and they violate the perturbative partial wave unitarity at the energy scale $4\pi M_{W}/g\sim1.5$~TeV\cite{higgs}.
A similar unitarity bound can be derived for massive fermion scattering amplitudes $f\bar{f}\to W^{+}_{L}W^{-}_{L}$. The unitarity bounds are inverse proportional to corresponding fermion masses  as $16\pi /\sqrt{2} G_{F} M_{f}\xi$  where $\xi = 1$ for leptons and $\xi = \sqrt{3}$ for quarks \cite{ac}. The strongest bound from the heaviest fermion top quark is about 3.5~TeV, much higher than the bound obtained from $WW$ scattering \cite{top}.
Given these arguments, even though the sector that 
give rise to SM fermion masses must break EWSB, 
it is not necessarily the dominant source 
for generating weak gauge boson masses
and in this paper, we investigate the implications
of such scenario in low energy measurements 
and collider physics.   

The models of EWSB are constrained by various
precision measurements. Among them, the $S$ and $T$ parameters are directly related to the properties of weak gauge bosons \cite{st}.
Flavor changing neutral current (FCNC) effects, on the other hand, 
directly constrain the mechanism of the SM fermion mass generation.
For instance, in extended TechniColor (ETC)
models \cite{etc}, in order to generate top quark mass,
the ETC scale must be low. However, such low ETC 
scale usually results in large FCNC effects since 
the ETC sector also couples to other fermions. 
Even for scenarios with SM fermion mass generation
through scalars, Glashow and Weinberg \cite{glashowweinberg} 
identified many of extended-scalar models produce
unacceptable FCNC even at the tree level.
The Glashow-Weinberg criterion for minimal flavor violation (MFV) 
has only two categories. One  case is where 
only one scalar doublet couples to the u-type quarks, and 
another scalar doublet couples to the d-type quarks.
The other case is that full $[U(3)]^{5}$ flavor symmetry of 
the standard model is broken by the Yukawa couplings 
of a single scalar doublet. In both cases, neutral scalars 
automatically have diagonal Yukawa couplings 
in the basis in which the quark mass matrix is diagonal.

Therefore, we keep the assumption that the SM fermion
masses are due to their coupling to  one doublet scalar
as the simplest scenario without large flavor violation at tree level. 
However, as mentioned, we study the case where 
the scalar is not solely responsible for the weak gauge boson 
masses. Given the additional source for EWSB, one general consequence of this scenario is that the Yukawa coupling is enhanced. 
To illustrate this feature, we use the bosonic Technicolor model \cite{bt1,bt2,bt3,bt4,bt5,bt6,bt7,bt8,bt9,carone}
as an example to study phenomenological implications of 
 scenarios where EWSB and SM fermion mass generation
come from two sources.

In the second section, we briefly discuss the spectrum of 
bosonic Technicolor model. 
There exist additional charged scalar state
which also couples to SM fermions with similarly enhanced couplings.  These enhanced Yukawa couplings 
may then result in large FCNC mediated by the charged scalar 
at loop level. In the third section, we use the FCNC especially
the neutral meson mixing to constrain the Yukawa coupling.
We then discuss the possible implications in the Higgs searches 
at the colliders in the allowed parameter region and 
conclude in the last section.

\section{Theory Realization}

Since SM fermion masses break the electroweak symmetry, 
the fermion mass generation sector  inevitably contributes to
weak gauge boson masses. The other sector that is responsible
for weak gauge boson masses is then fermiophobic \cite{Akeroyd:1995hg}.  
Fermiophobic can be achieved through either SM gauge symmetry
or additional symmetry. 

If the $SU(2)_{L}\times U(1)_{Y}$ symmetry is broken 
by the {\it vev} of a Higgs other than the doublet $(2,1)_{1}$ \cite{tripletlhc}, the SM gauge symmetries automatically 
forbid the coupling between such Higgs and the SM fermions
thus fermiophobic. 
On the other hand,  the non-doublet Higgs in general breaks the
$SU(2)_{L+R}$ custodial symmetry and generates unacceptable
contribution to $\Delta T$ so the {\it vev}s are usually constrained to
be smaller than $\cal O$(1~GeV).  One exception is the 
Georgi-Machacek model \cite{triplet} where two $SU(2)$ triplet 
Higgs are introduced and the $\Delta T$ is under control even with
larger {\it vev}. Such fermiophobic scalar $\phi$ has very unique
phenomenology since they can only be produced 
via $W\phi$, $Z\phi$ associate production and WBF. For 
light scalars $m_{\phi}< 130$~GeV, with the absence of $\phi b\bar{b}$ coupling, 
$\phi\to \gamma\gamma$ becomes the dominate decay channel \cite{ Akeroyd:1995hg,fermiophobichiggs}.
The heavy scalar decay would behave similarly to the SM-like Higgs except 
that there is no $t\bar{t}$ mode.  In addition, the $gg$ fusion production 
is absent and the search of such heavy fermiophobic Higgs is through $WWW$ channel as \cite{wuwei}.

Another realization is the Bosonic Technicolor
theory (BTC).  In this case, electroweak symmetry is broken by
 the condensation of
techni-quarks and the $WW$ scattering amplitude is unitarized by
 both techni-pion $\pi_{TC}$ and  techni-rho $\rho_{TC}$.
By assignment, the strongly interacting sector does not couple to
SM fermions. 
There exists a scalar doublet in the theory which does not develop {\it vev} but the scalar couples to both SM fermions and the techni-fermions.
The SM fermion masses arise via techni-fermions confinement 
and the strength of the Yukawa couplings still determines the mass of such fermion. 

In both fermiophobic Higgs models and the BTC model, the Yukawa couplings between the Higgs-like scalar and the SM fermions   
are in general enhanced  when
the EWSB is dominant due to a fermiophobic source.  In this paper, we use the BTC model to illustrate the general feature of enhanced
Yukawa couplings and discuss its phenomenology implications.

Using the non-linear representation, one can define $\Sigma$ in the 
electroweak chiral Lagrangian as in \cite{carone}
\be\label{eq:sigdef}
\Sigma = \exp(2 i \Pi/f), \,\,\,\,\,  \Pi = \left( \begin{array}{cc} \pi^0/2 & \pi^+/\sqrt{2} \\ \pi^-/\sqrt{2} & -\pi^0/2
\end{array}\right) \, ,
\ee
where $\Pi$ represents an isotriplet of techni-pions and $f$ is their decay constant. 
The theory also contains a scalar doublet $\Phi$ that couples to the SM fermions. 
\be
\Phi = \lf( \begin{array}{cc} \overline{\phi^0} & \phi^+ \\ -\phi^- & \phi^0 \end{array} \rh).
\ee
$\Phi$  mixes with the techni-pion state through Yukawa couplings 
to the techni-fermions. For convenience, one can rewrite the $\Phi$ field in
the non-linear form. By expanding around the true vacuum, one obtains
\be \label{eq:pisigma}
\Phi = \frac{\sigma+f'}{\sqrt{2}}\Sigma', \,\,\,\,\,\Sigma' = \exp(2 i \Pi'/f')\,,
\ee
where $f'$ is the {\it vev} of $\phi$ and $\Pi'$ represents its isotriplet components.    

The kinetic terms for the $\Phi$ and $\Sigma$ fields are 
\be \label{eq:chirallagrangian}
\mathcal L_{KE} = \frac{1}{2}\partial_\mu\sigma\partial^\mu\sigma
+\frac{f^2}{4}\textrm{Tr}(D_\mu\Sigma^\dagger D^\mu\Sigma)
+\frac{(\sigma+f')^2}{4}\textrm{Tr}(D_\mu\Sigma'^\dagger D^\mu\Sigma'),
\ee
where the covariant derivative is given by
\be
D^\mu\Sigma = \partial^\mu\Sigma-igW_a^\mu\frac{\tau^a}{2}\Sigma+ig'B^\mu\Sigma\frac{\tau^3}{2}.
\ee
For a specific linear combination of the pion fields:
\be \label{eq:absorbedmixing}
\pi_a = \frac{f\,\Pi+f'\,\Pi'}{\sqrt{f^2+f'^2}}.
\ee
there exist quadratic terms that mix the gauge fields with derivatives.
Such states are unphysical and can be gauged away.
The physical state $\pi_{p}$ then arises from the orthogonal linear combination,
\be \label{eq:physicalmixing}
\pi_p = \frac{-f'\,\Pi+f\,\Pi'}{\sqrt{f^2+f'^2}}\, .
\ee
The physical pion mass can be obtained by the potential term involving 
coupling between $\Phi$ and $\Sigma$ which originates from the techni-fermion
Yukawa coupling and techni-fermion condensation. The mass of physical pion
is then very model dependent. 
In unitary gauge, weak gauge boson masses arise from the remaining quadratic terms
\be
m_W^2 = \frac{1}{4}g^2v^2,\,\,\,\,\,\,\,\,\,\,  m_Z^2=\frac{1}{4}(g^2+g'^2)v^2,
\ee
where $v$ is the electroweak scale as $v \equiv \sqrt{f^2+f'^2} = 246 \textrm{ GeV}$.

For simplicity, we define a mixing angle in the following discussion 
\beq
\sin\theta = {f^{'}\over v}, \cos\theta = {f\over v}~.
\eeq
As long as the SM fermions get masses in the theory, $f^{'}$ is non-zero.
Therefore, $\cos\theta$ never reaches 1.   
In the limit of $\sin\theta =1$, the $\sigma$ field corresponds to the SM Higgs.
For other region, the $\sigma$ field behaves like  the SM Higgs but with different couplings. 
Expanding the third term of Eq.~(\ref{eq:chirallagrangian}), we find that
the coupling between $\sigma$ and the gauge bosons is given by
\be\label{eq:siggb}
\mathcal L_{\sigma WZ} = 2\sin\theta\f{m_W^2}{v} \sigma W^{+\mu}W_\mu^- +\sin\theta \f{m_Z^2}{v} \sigma Z^\mu Z_\mu \, ,
\ee
which is reduced by a factor of $\sin\theta$ compared to the result in the Standard Model. 

The couplings of the $\Phi$ field to the quarks is given by regular Yukawa coupling.
The coupling of the $\sigma$ field to fermions is given by
\be\label{eq:sff}
\mathcal L_{\sigma \bar f f} =  - \sum_{\textrm{fermions}} \f{1}{\sin\theta} \f{m_f}{v} \sigma \bar f f \, .
\ee
The $\sigma$ field is mostly $\Phi$-like. 
The $\sigma$ Yukawa coupling is larger than the SM Yukawa coupling
by a factor of  $1/\sin\theta=v/f'$. Similarly, the charged physical pion which is a mixing
state of $\phi^{\pm}$ and $\pi^{\pm}$ can also couple to the SM fermions.
However, given the physical charged pion state is a mixture of the
techni-pion and charged $\Phi$, the coupling is slightly different from 
the $\sigma$ by a factor of $\cos\theta=f/v$ and the couplings defer
$m_{f}/v$ by $\cos\theta/\sin\theta=\cot\theta$
\be
\mathcal L_{\pi^{\pm}_{p}}=-i\cot\theta {m_{u_{i}}\over v}\pi^{-}_p\bar{d}_{iL} u_{iR}-i\cot\theta{m_{d_{j}}\over v} \pi^+_p\bar{u}^{i}_{L}V^{ij}_{CKM} d^{j}_{R}+h.c.
\ee
Here, the index $i,j$ stands for flavor.
In the limit of $\sin\theta \rightarrow 1$, the charged degree of
freedom basically disappears and these couplings vanish. 

\section{Flavor constraints from enhanced Yukawa coupling}

The tree level FCNC mediated by the neutral scalar can be avoided
when the quark mass matrices and the Yukawa couplings
are diagonalized simultaneously \cite{glashowweinberg}. 
However, there exists the charged scalar $\pi^{+}_{p}$ that
couples to the SM fermions and the couplings are flavor violating. 
In addition, the enhanced couplings of $1/\sin\theta$  made
the constraints more severe. 
In this session, we examine the bounds 
on the Yukawa coupling due to FCNC processes involving the
charged scalar. 

At tree-level, the charged scalar can contribute to flavor
violating rare decays like $B_{u}\to \tau \nu_{\tau}$, $B\to D\tau\nu_{\tau}$. 
The SM prediction for the $\text{BR}(B_{u}\to \tau\nu_{\tau})_\text{SM}=(0.95\pm0.27)\times 10^{-4}$ while the current experimental value is
$\text{BR}(B_{u}\to \tau\nu_{\tau})_\text{exp}=(1.65\pm0.34)\times 10^{-4}$.
The $W^{\pm}$-mediated SM contribution is suppressed due to helicity suppression. Even though the charged scalar may be much heavier,
the no-helicity-suppression made the scalar-mediated contribution
comparable to the SM contribution. 
The decay amplitude is proportional to 
\beq
V_{ub} {m_{b}\over f^{'}}{m_{\tau}\over f^{'}} {f^{2}\over v^{2}}
\eeq
In the type-II 2HDM, a similar contribution mediated by the charged
Higgs only becomes relevant when the $\tan\beta \gtrsim 10$ for very
light charged Higgs of $M_{H^{+}}=100$~GeV and the bound 
for charged Higgs of 700~GeV is $\tan\beta < 60$ \cite{2hdmyukawa}. If one translates
such bound into the current model, the top quark Yukawa coupling
is greater than $4\pi$ and become strongly coupled. 
Therefore, in the range that we are interested in, the bound 
from $B_{u}\to \tau\nu_{\tau}$ is irrelevant. 

Recent search on $\mu\to e\gamma$ at MEG experiment will soon 
reach BR$(\mu\to e\gamma)\simeq 1\times 10^{-13}$. 
The one loop contribution from charged state to $\mu\to e\gamma$
is suppressed by the small lepton masses and additional 
helicity-flip. However, the usual largest contribution in 
Higgs mediated $\mu\to e \gamma$ is the Barr-Zee two-loop 
effects involving the charged scalar coupling to a top-bottom loop.
The two-loop involving $Z$ and neutral scalar does not 
exist since there is no tree-level flavor violating vertices in
the neutral scalar coupling to the SM fermions. 
But, again, similar study in type-II 2HDM has shown the contribution
only reach the next sensitivity for extremely large $\tan\beta$ of 60 \cite{hisano}. 
We don't expect the current model receive constraints from the lepton
flavor violating experiments like $\mu\to e\gamma$. 

Given the large top quark mass, the most serve constraints 
come from neutral $B_{d}$-mixing or $B_{s}$-mixing 
since these processes do not have suppression from $m_{b}$
or small lepton masses. 
The charged Higgs mediated $B_{d}$-mixing had been calculated 
for 2HDM \cite{bdmixing} 
\beq
\Delta M_{B_{d}}=\frac{G^{2}_{F}m^{2}_{t}f^{2}_{B_{d}}\hat{B}_{d}M_{B}\mid V^{*}_{td}V_{tb}\mid^{2}\eta_{b}}{24\pi^{2}}[I_{WW}(y^{W})+I_{W\Pi}(y^{W},y^{\Pi},x)+I_{\Pi\Pi}(y^{\Pi})],
\eeq 
where 
\beq
y_{W}={m^{2}_{t}\over m^{2}_{W}},~~y_{\Pi}={m^{2}_{t}\over m^{2}_{\pi^{+}_{p}}},~~x={m^{2}_{\pi^{+}_{p}}\over m^{2}_{W}}~,
\eeq
and
\begin{equation}
\begin{aligned}
I_{WW}&=1+\frac{9}{1-y^W}-\frac{6}{(1-y^W)^2}-\frac{6}{y^W}\left(\frac{y^W}{1-y^W}\right)^3\ln y^W\\
I_{W\Pi}&=\lambda_{tt}^2\, y^{\Pi}\left[\frac{(2x-8)\ln y^{\Pi}}{(1-x)(1-y^{\Pi})^2}+\frac{6x\ln y^W}{(1-x)(1-y^W)^2}-\frac{8-2y^W}{(1-y^W)(1-y^{\Pi})}\right]\\
I_{\Pi\Pi}&=\lambda_{tt}^4\, y^{\Pi}\left[\frac{1+y^{\Pi}}{(1-y^{\Pi})^2}+\frac{2y^{\Pi}\ln y^{\Pi}}{(1-y^{\Pi})^3} \right].
\end{aligned}
\end{equation}

The QCD running of the Wilson coefficient is given by $\eta_b=0.552$ \cite{buras}.
The non-perturbative decay constant $f_{B_d}$ and the bag parameter $\hat{B}_d$ from lattice QCD calculation is $f_{B_d}\hat{B}_d^{1/2}=216\pm 15$~MeV \cite{lattice} and
we take the conservative value as $f_{B_d}\hat{B}_d^{1/2}=201$~MeV.
Since the theoretical uncertainty from $f_{B}\hat{B}^{1/2}$ dominates experimental uncertainties and they are correlated in $B_{s}$ and $B_{d}$ decays, we 
use only $\Delta M_{B_d}$ here which has the smallest total uncertainty and
the current experiment measurement for $B_{d}$ mixing is
\beq
\Delta M_{B_{d}} = ( 507 \pm 4 ) \times 10^{9}  \hbar \text{s}^{-1} ~.
\eeq

\begin{figure}
\includegraphics[scale=1,width=7cm]{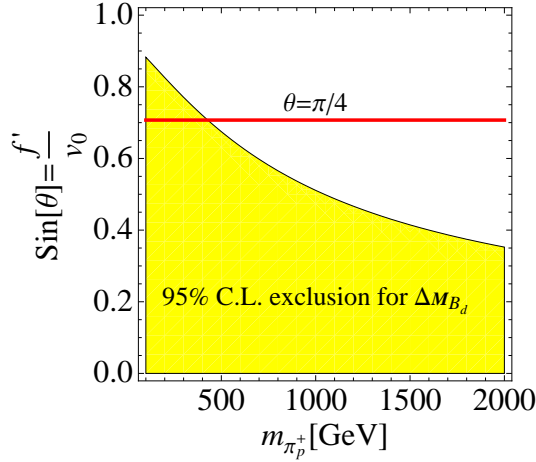}
\caption{Constraints on the $f^{'}/v$ and charged scalar mass $m_{\pi^{+}_{p}}$ from $\Delta M_{B_{d}}$. The yellow region is excluded at 95\% C.L. } 
\label{bdmixing_fig}
\end{figure}

Figure \ref{bdmixing_fig} shows the correlated constraints on the mixing $\sin\theta$ ($f^{'}/v$) and charged scalar mass $m_{\pi^{+}_{p}}$ from $\Delta M_{B_{d}}$. The mixing $\sin\theta$ represents the ratio of
top quark mass effects in the total EWSB. In the limit of $\sin\theta=1$, the $f^{'}=v$
and the charged scalar state disappears and the $\sigma$ corresponds
to the SM Higgs boson. 

The correlation showed in Fig. \ref{bdmixing_fig} gives the constraints on the charged scalar
masses and the charged scalar coupling to the SM fermions.   
The mixing $\theta=\pi/4$ corresponds to where the $f^{'}$ is 50\% of the EWSB and the $\sigma$ coupling to SM fermions have been enhanced by a factor of 2.
As long as the charged scalar mass is greater than 500~GeV, the scenario
is not excluded by the $\Delta M _{B_{d}}$. 
In the following session, we use the factor of 2 enhancement 
as a benchmark point to illustrate its strong implications over the Higgs phenomenology. 

\section{Phenomenological Implications}

In the BTC models, 
the couplings of the lightest neutral scalar $\sigma$ to 
weak gauge bosons are suppressed by $\sin\theta$ while to the SM fermions are enhanced by $1/\sin\theta$ in comparison to the SM Higgs boson. 
This feature significantly changes the Higgs phenomenology. 
An immediate consequence in the $\sigma$ production is that
the gluon fusion production rate is enlarged  while at the same 
time, the $W\sigma$ associated production rate as well as
the weak boson fusion production rate are both reduced.
On the other hand, the $\sigma$ decay BR is also modified.
In this section, we use the benchmark point
of $\sin\theta=\pi/4$ in BTC model to illustrate how the 
Higgs phenomenology changes accordingly in this scenario. 

The most important channel for light SM Higgs discovery is the
di-photon search. Similarly, the $\sigma\to \gamma\gamma$ 
has both $WW$ and fermion loop contribution while
the fermion loop contribution is enhanced and $WW$ loop is reduced. 
The $\sigma\to \gamma\gamma$ decay partial width is given by 
\beq
\Gamma(\sigma\to \gamma\gamma) = \frac{G_{\mu}\alpha^{2} M^{3}_{H}}{128\sqrt{2}\pi^{3}}\mid \sum_{f}N_{c} Q^{2}_{f}A^{H}_{1/2}(\tau_{f}) /\sin\theta+A^{H}_{1}(\tau_{W})\sin\theta\mid^{2}
\eeq
where
\bea
A^{H}_{1/2} (\tau) & = & 2[\tau+(\tau-1)f(\tau)]\tau^{-2}\nonumber\\
A^{H}_{1}(\tau) & = & -[2\tau^{2}+3\tau+3(2\tau-1)f(\tau)]\tau^{-2} 
\eea
and the function $f(\tau)$ is defined as
\beq
f(\tau)=\left \{
\begin{array} {rl}
 \arcsin^{2} \sqrt{\tau}; & \mbox{ $\tau \le$ 1} \\
 -{1\over 4} \left[ \log\frac{1+\sqrt{1-\tau^{-1}}}{1-\sqrt{1-\tau^{-1}}}\right]^{2}; &\mbox{ $\tau >$ 1}
\end{array}\right.
\eeq
One interesting feature is that the $W$-loop and heavy quark loop interfere 
destructively. In the case of SM Higgs boson, the contribution is dominated
by the $W$-loop contribution. However, with the reduced coupling in $\sigma WW$
and the enhanced coupling in $\sigma f\bar{f}$, the destructive interference
effect become significant in the di-photon channel and makes the 
$\Gamma(\sigma\to \gamma\gamma)$ partial width much smaller than the
SM Higgs for the corresponding masses. 
\begin{figure}
\includegraphics[scale=1,width=7cm]{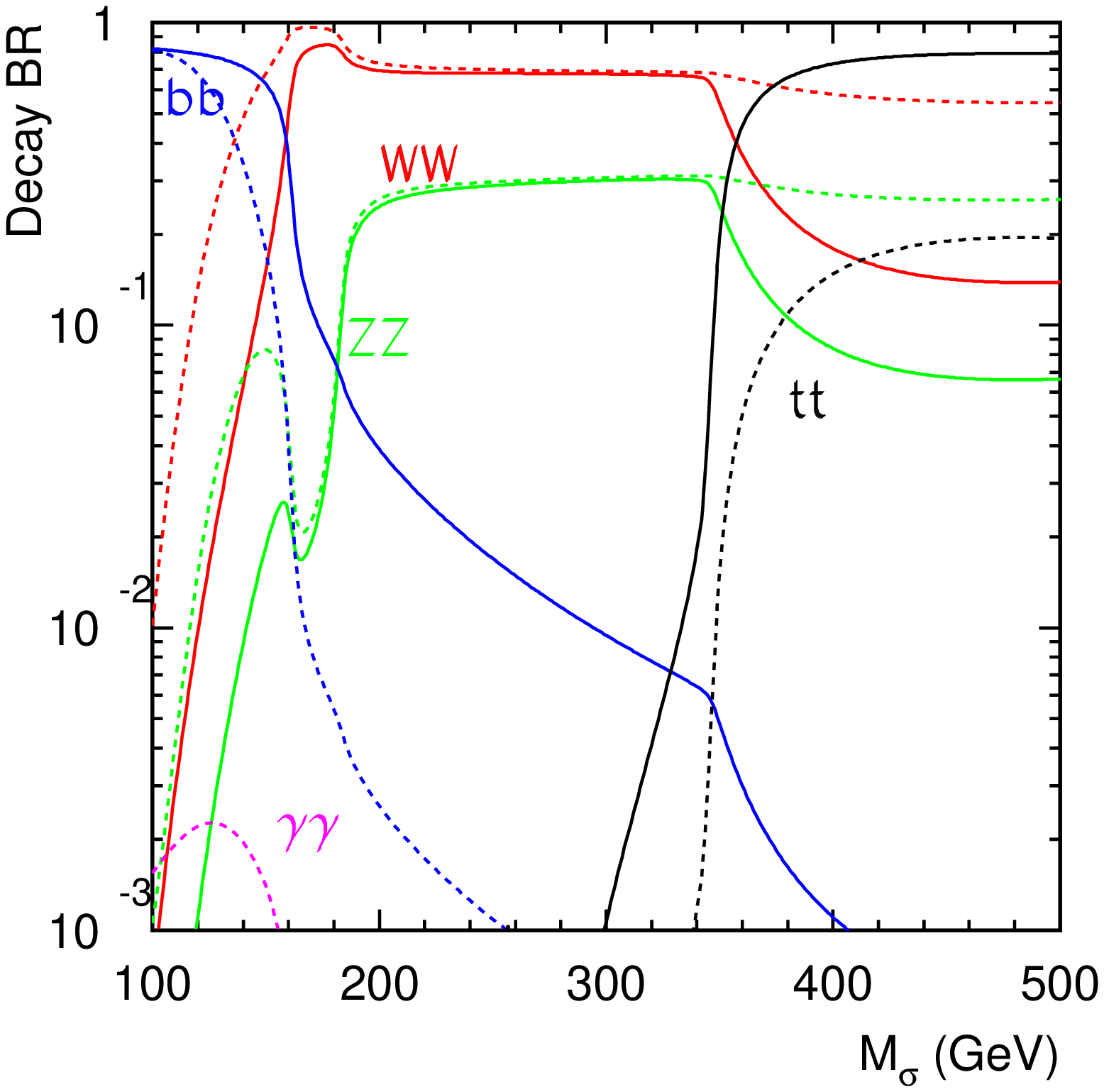}
\includegraphics[scale=1,width=7cm]{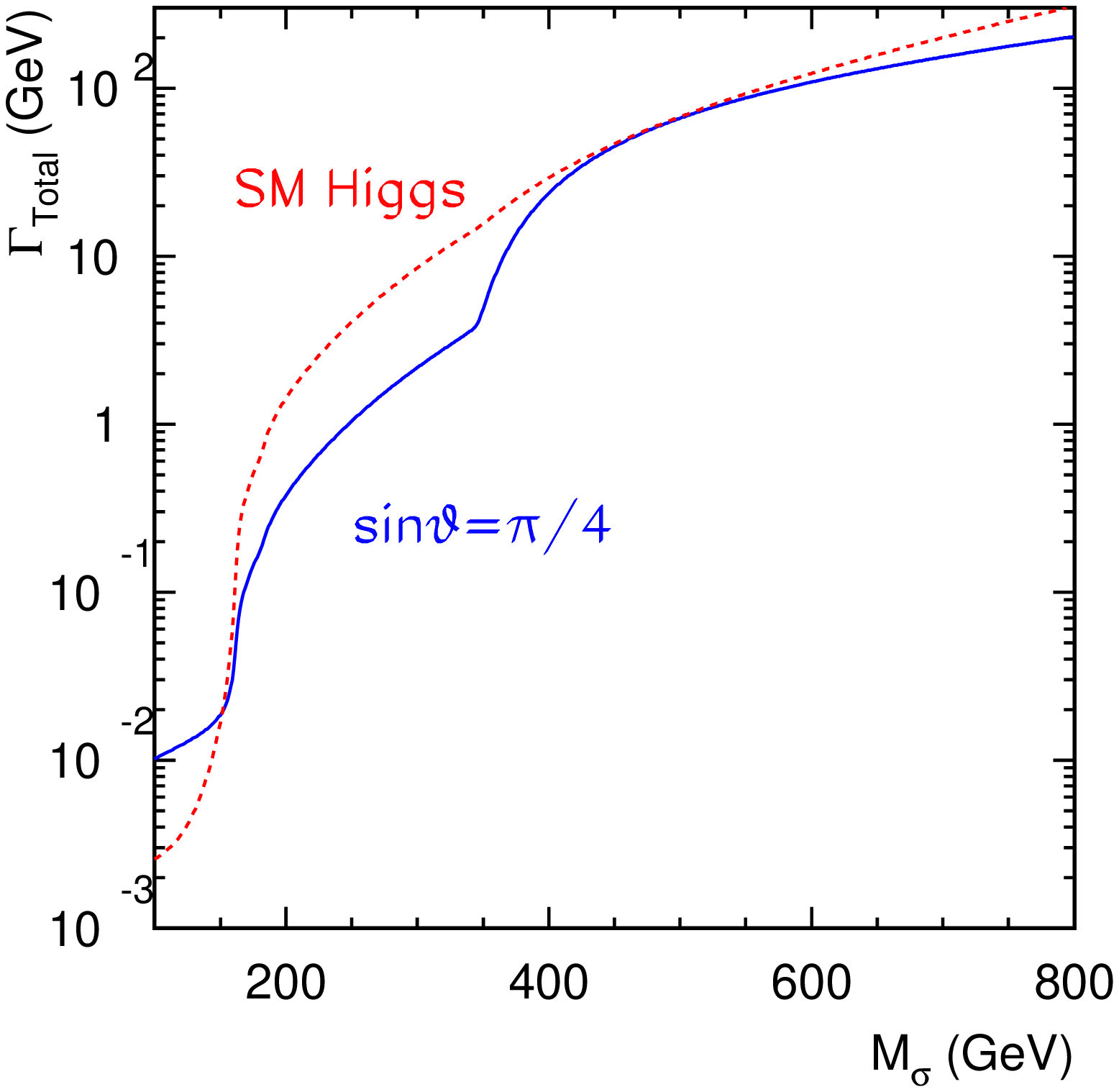}
\caption{(a)Decay BR of $\sigma$ (solid lines); (b) $\sigma$ total width (solid line), both
in the case of $\sin\theta=\pi/4$ $(f'=f)$, in comparison with the SM Higgs boson (dashed lines) . 
}
\label{higgsdecay}
\end{figure}
Figure \ref{higgsdecay} shows (a) the decay BR of $\sigma$  to various 
important searching modes and (b) the total width of $\sigma$ 
in the case of mixing $\sin\theta=\pi/4$.
The partial width of $\Gamma(\sigma\to \gamma\gamma)$ for $M_{\sigma}=120$~GeV is only $\cal O$(0.1~KeV) which is one order of magnitude smaller than 
the corresponding partial width for SM Higgs boson. In this mass region,
$b\bar{b}$ mode dominates the decay and the partial width is enhanced
by factor of 4. Consequently, the BR$(\sigma\to \gamma\gamma)$ is reduced
by two order of magnitude. 
The di-photon mode is essentially negligible even
with the enhancement in top Yukawa coupling of $gg$ fusion.
The $\sigma$ production via gluon fusion with $\sigma\to b\bar{b}$ 
decay would encounter tremendous irreducible QCD $b\bar{b}$ background.
Another standard search through associate production $W\sigma$, $Z\sigma$ 
with with $\sigma\to b\bar{b}$ is suppressed by factor of $\sin^{2}\theta$. 
The light $\sigma$ search is then much more challenging than the SM Higgs boson. 

In the off-shell $W$ region, given the enhancement
in $b\bar{b}$ and reduce in $WW^{*}$, $b\bar{b}$ dominates larger region than 
the SM Higgs. In Fig.\ref{higgsdecay}, $M_{\sigma}=140$~GeV, BR($\sigma\to WW^{*}$) is almost 8 times smaller than the corresponding SM value. Therefore, even
with the enhancement in gluon fusion production, the $WW^{*}$ expectation
in this scenario is smaller than the rate of the SM Higgs.

While for on-shell $WW$,$ZZ$ region, the $WW$ or $ZZ$ 
partial width is reduced but they are still much larger than the $b\bar{b}$. 
In the on-shell $WW$,$ZZ$ mass region, the $\sigma$ decay BR is still dominated 
by the weak gauge boson mode. Given the enhanced
top Yukawa coupling, the exclusion from Higgs direct search via $ZZ\to 4\ell$ 
is much more severe in this case. Only when the $\sigma$ mass is larger than the $2 m_{t}$ threshold, the enhanced Yukawa coupling then increases 
BR($\sigma\to t\bar{t}$) significantly and the $WW$,$ZZ$ decay BR
are reduced by factor of 4. The direct search bound for SM Higgs can then be applied 
here.

\section{Conclusions}
In this paper, we use the bosonic Technicolor (BTC) model to 
illustrate some general features of scenarios where 
weak gauge boson mass generation is 
dominated by a fermiophobic sector but the SM fermion mass generation
is still due to a light scalar doublet. In this case,
the Yukawa couplings between the light Higgs and the standard
model fermions are enhanced while the couplings to weak
gauge bosons are reduced.  
The remaining charged scalar state inevitably mediate flavor changing
neutral current processes with enhanced Yukawa
couplings. We find that $\Delta M_{B_{d}}$ is the most severe bound.
The correlation between the Yukawa coupling 
and charged scalar mass show that a factor of two 
enhancement in top Yukawa coupling is still allowed
for charged scalar heavier than 500~GeV. 
We use this benchmark point to study the phenomenology
of the light scalar. The destructive interference
between the top quark loop and the $W$-boson loop in the di-photon channel
becomes significant and makes the di-photon negligible. 
For the light scalar region, the search becomes much more 
challenging than the conventional SM Higgs boson. 
We also discuss the implication on the light scalar search
for other mass region and find for heavy scalar, the 
direct search bound is more severe.

\section*{Acknowledgement}
We would like to thank Guohuai Zhu for useful discussion. 
The work is supported by the Fundamental Research Funds for 
the Central Universities 2011QNA3017. ML is also supported in part by the National Science Foundation of China (10875103), National Basic Research Program of China (2010CB833000), and Zhejiang University Group Funding (2009QNA3015).

\end{document}